\documentclass[preprint,12pt]{elsarticle}

\usepackage{amssymb}

\journal{Biosystems}

\begin{document}

\begin{frontmatter}



\title{Stochastic Chemical Kinetics. Theory and (Mostly) Systems Biological Applications, P. Erdi, G. Lente. Springer (2014)}


\author{Enrico Bibbona }

\address{Dipartment of Mathematics ``G.Peano'', University of Torino\\
enrico.bibbona@unito.it}

\end{frontmatter}


The idea of modeling chemical reaction is rather old, but the topic has gained renewed interest in recent years. The newest experimental techniques make it possible to collect data from networks of reactions that take place at the level of a single cell and to compare experimental evidence with model predictions. Such an ``application-oriented'' interest motivated the authors to write the new book \cite{erdiNew}, which may serve both as an introduction to classical stochastic models, as well as a guide into the vast literature that appeared recently.
The book is highly readable and does not require mastery of mathematics at a highly specialized level: it rather addresses a wide audience including students and applied scientists.

One of the authors, P. Erdi, together with J. Toth, had already authored a book on \emph{Mathematical Modelling of Chemical Reactions} in 1989 \cite{erdiOld}. The two books are rather different, the elder is longer, more mathematically oriented and focused on the non-linear aspects of the theory.

The structure of the new book \cite{erdiNew} is as follows.
Chapter 1 shortly introduces the law of mass action and the deterministic models of chemical kinetics. Such models based on ordinary differential equations (ODEs) became popular in the 1960s-1970s among mathematicians and physicists since their non-linear character made them a natural field of application of mathematical ideas like (multi-)stationarity, bifurcation and deterministic chaos. A more detailed account of the theory can be found in the already mentioned elder book  \cite{erdiOld} and elsewhere (a detailed list of references is provided at the end of the chapter). Let me also mention a course that M. Feinberg, one of the major contributors to this topic, delivered at the Mathematics Research Center of the University of Wisconsin-Madison in 1979 (cf. \cite{deterministic1979}) which is a remarkable account of the state of the art at that time.

The stochastic models are presented at the end of the chapter together with an historical perspective. They entered the field more or less contemporarily to the deterministic counterparts. Again those models were introduced and discussed mainly between theoreticians, but they had an enormous success in addressing the inherent random character of the phenomenon. Stochastic models became popular tools in chemical physics and mathematical biology, maybe also because they have been formulated in terms of a class of random processes - Markov Chains - whose theory is accessible to the more applied scientist without a strong mathematical background.

Despite the apparent theoretical simplicity, however, in such models it is often very hard to calculate any quantity explicitly and in the case of large systems, even numerical solutions or direct simulations may be impractical. The numerical difficulties stimulated the development of new methods at the interface between mathematics, physics and computer science. A short account of such methodologies is presented in Chapter 2. The chapter begins by sketching a brief and gentle overview of some stochastic processes and their main properties giving references to the relevant literature, and continues by listing some methods of solutions of the master equations both in the transient and in the stationary regimen. The occurrence of multimodal behaviors is discussed.  Some algorithms for efficient simulations are also explained.
A distinguished feature of this book is to discuss rather systematically the applicability limits of deterministic kinetics and to explain which are the conditions in which only the stochastic approach is viable. Stochastic maps are also introduced with this purpose.

Chapter 3 is devoted to applications and it is very rich. It starts from some example phenomena where deterministic and stochastic modes are compared. Compartmental models are then treated in some detail showing how the methods introduced in Chapter 2 may prove useful in such context. The phenomenon of autocatalysis is then explained and the experimental observations are mentioned. Enzyme-catalyzed reactions are the next topic, the Michaelis-Menten equation and its solution are discussed and more complicated enzyme system are also sketched, with reference to experimental works where the phenomenon of extinction was observed. Section 6 of this chapter provides an alternative view of chemical systems: it is shown, also by many examples, how they can be interpreted as signal processing devices and an interesting application to the olfactory system is illustrated. Gene expression is a very popular topic which largely benefits from stochastic modelling and an account of such applications is also found in this chapter. The origins of molecular chirality are then addressed, starting from simple unrealistic models that can be analyzed analytically and concluding with the more complex Soai reaction that needs a clever combination of simulations and further different methods to be quantitatively explained. The chapter ends with two further topics: parameter estimation and stochastic resonance.

\medskip

The readership that would benefit by reading this book includes the following categories. Students of mathematics (both graduate and undergraduate) who want to see relevant applications in natural sciences (I have already suggested part of this book for a thesis at the bachelor level). Students in physics or theoretical and computational biology who want to see how the stochasticity plays a role in many examples of practical relevance. Researchers who need a recent survey of the topic.

This book has several pros:
\begin{itemize}
\item It is short and very well focused, not a single word has been wasted in useless discussion and the reader is quickly driven to what they need for their application.
\item A lot of care has been taken in order to make the book accessible to the broadest readership: neither a very strong mathematical background, nor a previous knowledge of system biology is required.
\item Both authors gave remarkable research contributions on the topic (cf. e.g. \cite{lenteA, lenteB,lenteC,lenteD,erdiA,erdiB,erdiC}) and their experience is clearly reflected in the choice of the topics and in how they are presented.
\item The book is very well documented and includes references to hundreds of research papers and to many books.
\end{itemize}
Every reader would like to see reflected in a book his personal experience and taste and I'm not an exception. A rigorous mathematical clarification of the relation between stochastic and deterministic models came with the work of T. Kurtz: he rigorously proved  (cf. \cite{kurtz1, kurtz4, kurtzBook} and the recent review \cite{kurtz5}) that  stochastic models converge to the deterministic ones as the size of the system gets large (as in mean field limits) and he also provided a diffusion approximation (often referred to under the name of Langevin equations, cf. \cite{kurtz2, kurtz3}) whose applicability has often been underestimated.
His profound contributions are just shortly mentioned in the book, and they could have deserved more space. A more detailed explanation of these beautiful and powerful concepts, however, would probably have raised the amount of mathematical details to be presented, and that probably explains why the authors made a different choice. Let me close the review by citing a few contributions to the topic that are simply too recent to have been included in the book but that I think (in some case I hope) the readership of the book could appreciate too: they are \cite{MR3178496,MR3059268, MR2288709,MR3199985, LNCS,TCS}.



\bibliographystyle{elsarticle-num} 
\bibliography{bibl}

\begin{thebibliography}{10}
\expandafter\ifx\csname url\endcsname\relax
  \def\url#1{\texttt{#1}}\fi
\expandafter\ifx\csname urlprefix\endcsname\relax\def\urlprefix{URL }\fi
\expandafter\ifx\csname href\endcsname\relax
  \def\href#1#2{#2} \def\path#1{#1}\fi

\bibitem{erdiNew}
P.~{\'E}rdi, G.~Lente, Stochastic Chemical Kinetics, Springer Series in
  Synergetics, Springer, 2014.

\bibitem{erdiOld}
P.~{\'E}rdi, Mathematical models of chemical reactions: theory and applications
  of deterministic and stochastic models, Manchester University Press, 1989.

\bibitem{deterministic1979}
M.~Feinberg, Lectures on reaction networks,
  \url{http://crnt.engineering.osu.edu/LecturesOnReactionNetworks}.

\bibitem{lenteA}
{\'E}.~D{\'o}ka, G.~Lente, Mechanism-based chemical understanding of chiral
  symmetry breaking in the soai reaction. a combined probabilistic and
  deterministic description of chemical reactions, Journal of the American
  Chemical Society 133~(44) (2011) 17878--17881.

\bibitem{lenteB}
G.~Lente, Homogeneous chiral autocatalysis: a simple, purely stochastic kinetic
  model, The Journal of Physical Chemistry A 108~(44) (2004) 9475--9478.

\bibitem{lenteC}
G.~Lente, Stochastic kinetic models of chiral autocatalysis: a general tool for
  the quantitative interpretation of total asymmetric synthesis, The Journal of
  Physical Chemistry A 109~(48) (2005) 11058--11063.

\bibitem{lenteD}
G.~Lente, Stochastic mapping of first order reaction networks: A systematic
  comparison of the stochastic and deterministic kinetic approaches, The
  Journal of chemical physics 137~(16) (2012) 164101.

\bibitem{erdiA}
P.~{\'E}rdi, J.~T{\'o}th, V.~Hars, Some kinds of exotic phenomena in chemical
  systems, in: Qualitative theory of differential equations (Szeged). Colloquia
  Mathematica Societatis J{\'a}nos Bolyai, Vol.~30, 1981, pp. 205--229.

\bibitem{erdiB}
T.~Sipos, J.~T{\'o}th, P.~{\'E}rdi, Stochastic simulation of chemical reaction
  by digital computer, i. the model, Reaction Kinetics and Catalysis Letters
  1~(1) (1974) 113--117.
\newblock \href {http://dx.doi.org/10.1007/BF02075130}
  {\path{doi:10.1007/BF02075130}}.

\bibitem{erdiC}
T.~Sipos, J.~T{\'o}th, P.~{\'E}rdi, Stochastic simulation of chemical reactions
  by digital computer, ii. applications, Reaction Kinetics and Catalysis
  Letters 1~(2) (1974) 209--213.
\newblock \href {http://dx.doi.org/10.1007/BF02067542}
  {\path{doi:10.1007/BF02067542}}.

\bibitem{kurtz1}
T.~G. Kurtz, Solutions of ordinary differential equations as limits of pure
  jump {M}arkov processes, Journal of Applied Probability 1~(7) (1970) 49--58.

\bibitem{kurtz4}
T.~G. Kurtz, The relationship between stochastic and deterministic models for
  chemical reactions, The Journal of Chemical Physics 57~(7) (1972) 2976--2978.

\bibitem{kurtzBook}
S.~N. Ethier, T.~G. Kurtz, Markov processes: characterization and convergence,
  Vol. 282, John Wiley \& Sons, 2009.

\bibitem{kurtz5}
D.~F. Anderson, T.~G. Kurtz, Continuous time markov chain models for chemical
  reaction networks, in: Design and analysis of biomolecular circuits,
  Springer, 2011, pp. 3--42.

\bibitem{kurtz2}
T.~G. Kurtz, Limit theorems and diffusion approximations for density dependent
  markov chains, in: Stochastic Systems: Modeling, Identification and
  Optimization, I, Springer, 1976, pp. 67--78.

\bibitem{kurtz3}
T.~G. Kurtz, Strong approximation theorems for density dependent markov chains,
  Stochastic Processes and Their Applications 6~(3) (1978) 223--240.

\bibitem{MR3178496}
H.-W. Kang, T.~G. Kurtz, L.~Popovic,
  \href{http://dx.doi.org/10.1214/13-AAP934}{Central limit theorems and
  diffusion approximations for multiscale {M}arkov chain models}, Ann. Appl.
  Probab. 24~(2) (2014) 721--759.
\newblock \href {http://dx.doi.org/10.1214/13-AAP934}
  {\path{doi:10.1214/13-AAP934}}.
\newline\urlprefix\url{http://dx.doi.org/10.1214/13-AAP934}

\bibitem{MR3059268}
H.-W. Kang, T.~G. Kurtz, \href{http://dx.doi.org/10.1214/12-AAP841}{Separation
  of time-scales and model reduction for stochastic reaction networks}, Ann.
  Appl. Probab. 23~(2) (2013) 529--583.
\newblock \href {http://dx.doi.org/10.1214/12-AAP841}
  {\path{doi:10.1214/12-AAP841}}.
\newline\urlprefix\url{http://dx.doi.org/10.1214/12-AAP841}

\bibitem{MR2288709}
K.~Ball, T.~G. Kurtz, L.~Popovic, G.~Rempala,
  \href{http://dx.doi.org/10.1214/105051606000000420}{Asymptotic analysis of
  multiscale approximations to reaction networks}, Ann. Appl. Probab. 16~(4)
  (2006) 1925--1961.
\newblock \href {http://dx.doi.org/10.1214/105051606000000420}
  {\path{doi:10.1214/105051606000000420}}.
\newline\urlprefix\url{http://dx.doi.org/10.1214/105051606000000420}

\bibitem{MR3199985}
J.~K. McSweeney, L.~Popovic,
  \href{http://dx.doi.org/10.1214/13-AAP946}{Stochastically-induced bistability
  in chemical reaction systems}, Ann. Appl. Probab. 24~(3) (2014) 1226--1268.
\newblock \href {http://dx.doi.org/10.1214/13-AAP946}
  {\path{doi:10.1214/13-AAP946}}.
\newline\urlprefix\url{http://dx.doi.org/10.1214/13-AAP946}

\bibitem{LNCS}
M.~Beccuti, E.~Bibbona, A.~Horvath, R.~Sirovich, A.~Angius, G.~Balbo, Analysis
  of petri net models through stochastic differential equations, in: G.~Ciardo,
  E.~Kindler (Eds.), Application and Theory of Petri Nets and Concurrency, Vol.
  8489 of Lecture Notes in Computer Science, Springer International Publishing,
  2014, pp. 273--293.

\bibitem{TCS}
A.~Angius, G.~Balbo, M.~Beccuti, E.~Bibbona, A.~Horvath, R.~Sirovich,
  Approximate analysis of biological systems by hybrid switching jump
  diffusion, arXiv preprint arXiv:1406.1352.

\end{thebibliography}
\end{document}